\documentclass[preprint,12pt]{elsarticle} 

\usepackage{lineno}
\modulolinenumbers[5]
\usepackage{color}
\usepackage[usenames,dvipsnames]{xcolor}
\usepackage{fullpage}
\usepackage{amsmath}
\usepackage{amssymb}
\usepackage{mathrsfs}
\usepackage{newfloat}
\usepackage{graphicx}
\usepackage{subfig}
\usepackage{multirow}
\usepackage{stackengine}
\usepackage{caption}
\usepackage{tikz}
\usepackage{float}
\usepackage{multicol}
\usepackage[inline]{enumitem}
\usetikzlibrary{arrows,positioning,calc}
\usepackage{pstool}
\usepackage{textcomp}
\usepackage{fixltx2e}
\usepackage[hidelinks]{hyperref}
\hypersetup{colorlinks,bookmarksopen,linkcolor=blue,citecolor=blue,urlcolor=blue}

\definecolor{myblue}{RGB}{0,115,189}

\DeclareFloatingEnvironment{algorithm}
\DeclareFloatingEnvironment{procedure}
\pgfdeclarelayer{bg} 
\pgfsetlayers{bg,main} 
\graphicspath{{figs/}}

\definecolor{g}{rgb}{0,0.6,0.3}
\definecolor{b}{rgb}{0,0.4,0.8}

\newcommand{\rev}[1]{{\color{black}#1}}

\journal{Mater. Des.}

\biboptions{authoryear}

\begin{document}

\begin{frontmatter}
\rev{
\title{
Experimental Full-field Analysis of Size Effects in Miniaturized Cellular Elastomeric Metamaterials\tnoteref{mytitlenote}
}
\tnotetext[mytitlenote]{The post-print version of this article is published in \emph{Mater. Des.}, \href{https://doi.org/10.1016/j.matdes.2020.108684}{10.1016/j.matdes.2020.108684}}
}
\author[TUe]{S.~Maraghechi} 
\ead{S.Maraghechi@tue.nl}

\author[TUe]{{J.P.M.~Hoefnagels}\corref{correspondingauthor}}
\ead{J.P.M.Hoefnagels@tue.nl}

\author[TUe]{R.H.J.~Peerlings}
\ead{R.H.J.Peerlings@tue.nl}

\author[TUe]{O.~Roko\v{s}}
\ead{O.Rokos@tue.nl}

\author[TUe]{M.G.D.~Geers}
\ead{M.G.D.Geers@tue.nl}

\address[TUe]{Mechanics of Materials, Department of Mechanical Engineering, Eindhoven University of Technology, P.O.~Box~513, 5600~MB~Eindhoven, The~Netherlands}
\cortext[correspondingauthor]{Corresponding author.}

\begin{abstract}

\rev{Cellular elastomeric metamaterials are interesting for various applications, e.g.\ soft robotics, as they may exhibit multiple microstructural pattern transformations, each with its characteristic mechanical behavior. 
Numerical literature studies revealed that pattern formation is restricted in (thick) boundary layers causing significant mechanical size effects. 
This paper aims to experimentally validate these findings on miniaturized specimens, relevant for real applications, and to investigate the effect of increased geometrical and material imperfections resulting from specimen miniaturization. 
To this end, miniaturized cellular metamaterial specimens are manufactured with different scale ratios, subjected to in-situ micro-compression tests combined with digital image correlation yielding full-field kinematics, and compared to complementary numerical simulations.
The specimens’ global behavior agrees well with the numerical predictions, in terms of pre-buckling stiffness, buckling strain and post-buckling stress. 
Their local behavior, i.e.\ pattern transformation and boundary layer formation, is also consistent between experiments and simulations. 
Comparison of these results with idealized numerical studies from literature reveals the influence of the boundary conditions in real cellular metamaterial applications, e.g. lateral confinement, on the mechanical response in terms of size effects and boundary layer formation.}


\end{abstract}

\begin{keyword}
Cellular Elastomeric Materials   \sep  Metamaterials  \sep  Digital Image Correlation  \sep Size Effects \sep \textit{In-situ} testing  \sep Microstructural Buckling
\end{keyword}

\end{frontmatter}



%
%
\section{Introduction}
\label{sec4:intro}

\begin{figure}
	\centering
	\subfloat[Undeformed microstructure]{\includegraphics[width=.3\textwidth]{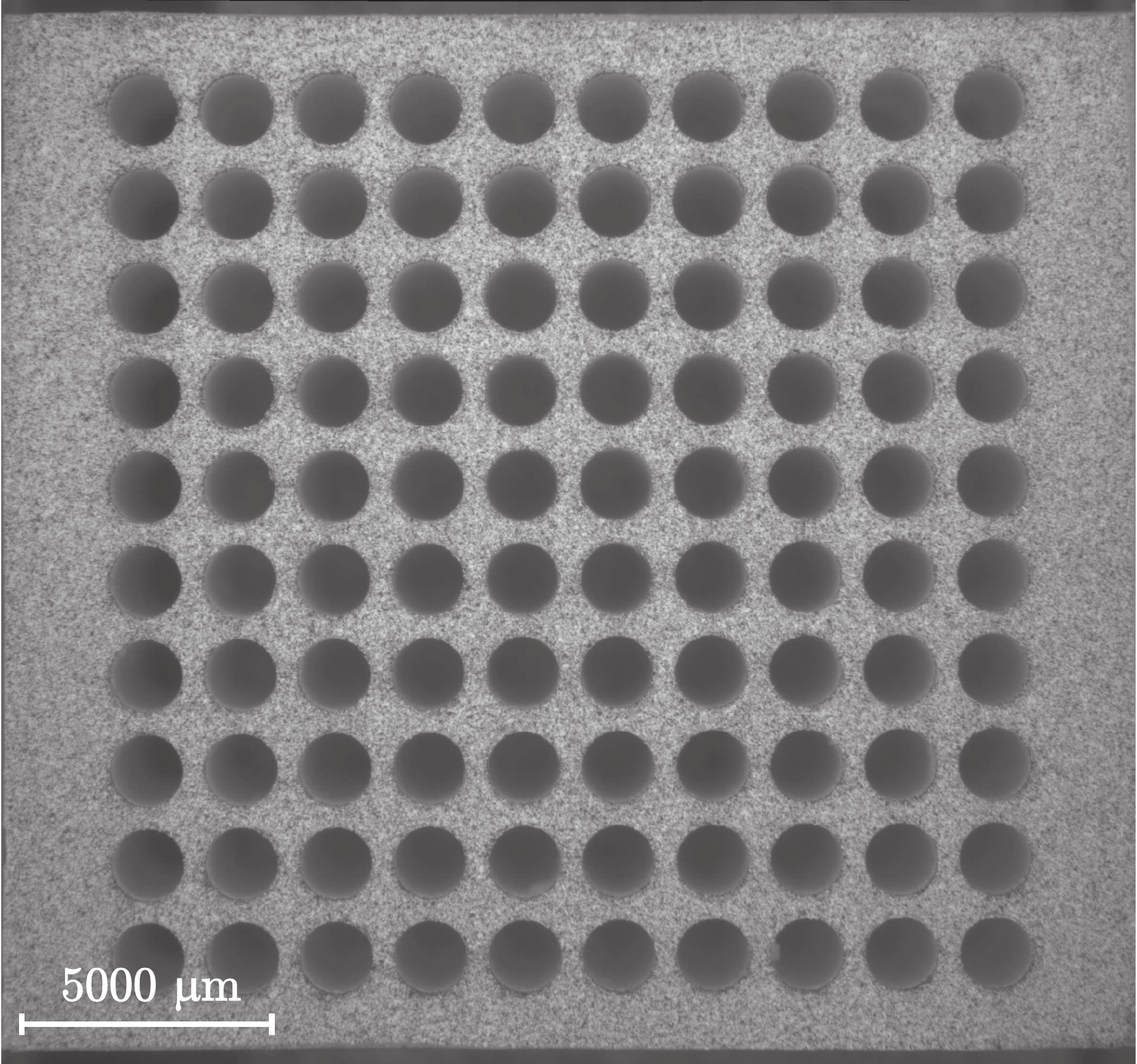}\label{fig4:undef}}\hspace{1.0em}
	\subfloat[Deformed microstructure]{\includegraphics[width=.3\textwidth]{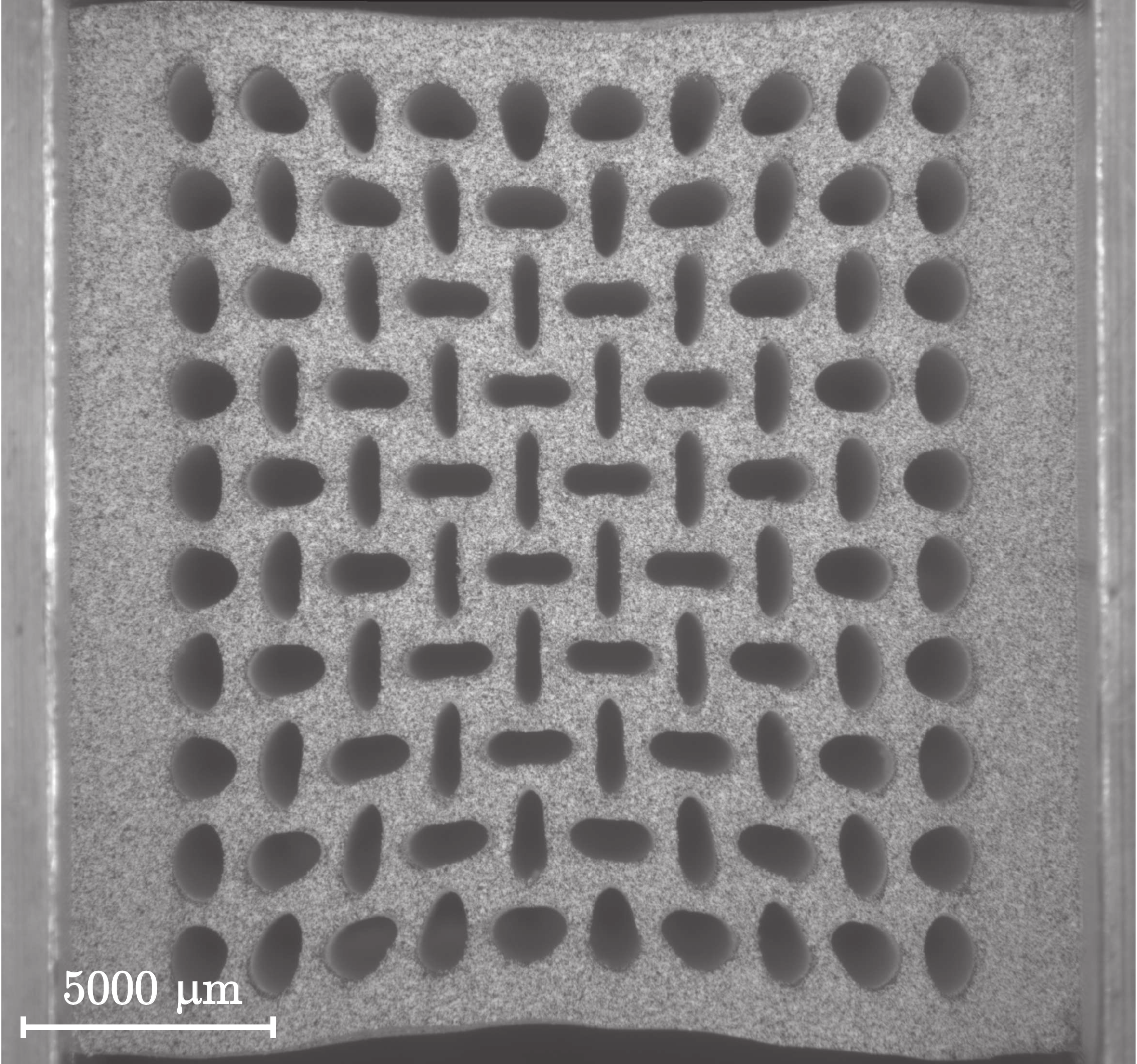}\label{fig4:def}}
	\caption{\rev{A cellular elastomeric metamaterial \protect\subref{fig4:undef} before, and \protect\subref{fig4:def} after the onset of buckling in the microstructure due to compression in the horizontal direction, leading to the emergence of an anti-symmetric pattern.}}
	\label{fig4:pattern}
\end{figure}

Cellular soft materials have proven to be useful in a variety of applications, and thus have attracted many studies in the past decade \citep{Yu2018,Wu2019,Jia2019}.
\rev{The behaviour of such materials significantly depends on their microstructural geometry, where buckling of the microstructure entails a non-local effective behaviour, exhibiting one or multiple emerging patterns, e.g.\ the pattern of Fig.~\ref{fig4:def} showing the deformed state with respect to the reference state of Fig.~\ref{fig4:undef}.}
Since these materials typically reveal a distinctly different mechanical behaviour beyond their buckling point compared to before, e.g.\ a transition from non-auxetic to auxetic behaviour \citep{Bertoldi2010,Yang2018,Mizzi2018}, they are commonly categorized as metamaterials.

\rev{
Whereas the majority of detailed experimental studies on cellular elastomeric metamaterials has been conducted on large specimens, most of the real applications of such metamaterials require a smaller characteristic size of the microstructure.
See e.g.\ \citep{Yang2015,Mirzaali2018a}, where small structures for handling delicate objects are envisioned as a promising application in soft robotics.
In such examples, the overall size of the structure is typically small as well, i.e.\ consisting of a few microstructural unit cells, making its mechanical response sensitive to the applied boundary conditions.}
In cellular elastomeric metamaterials, the pattern transformation is strongly affected by the boundary conditions that may locally confine the patterning.
Kinematic restrictions lead to the formation of boundary layers resulting in a significant dependence of the effective mechanical response on the ratio of the specimen size to hole size, i.e. the so-called ``size effect''. 
In a recent study, \citet{Ameen2018} numerically investigated the size effect in cellular elastomers.
They showed a considerable influence of size on the global behaviour of such materials and a characteristic boundary layer size in specimens with scale ratios ranging from 4 to 128, where scale ratio is the ratio of the specimen length to the length of a unit cell of the microstructure.
\rev{They reported an increasing trend in the boundary layer thickness with an asymptote of approximately three unit cells and an asymptotic trend of global stress with more than 40\% decrease over the range of scale ratios.
This trend is most considerable for smaller scale ratios, i.e.\ 29\% decrease for scale ratio of 12.}

\rev{
The goal of this paper is to experimentally verify the numerically observed size effects for miniaturized cellular metamaterial specimens suitable for realistic applications.
The challenges in manufacturing of miniaturized metamaterial specimens lead to imperfections in the material, thus making it important to verify the numerically observed size effects on such specimens and not only on large scale ones, as typically found in the literature.

For instance, \cite{coulais2018}  investigated size effects in cellular metamaterials under inhomogeneous loadings. 
\cite{Zhang2019} studied the size-dependence of soliton number and wavelength in metamaterials.
\cite{dunn2020} investigated the size effects in 3D metamaterials under bending and torsion.
However, all of these studies focus on large specimens, typically with unit cells of centimetre scale.
There are some studies in the literature on mico-scale metamaterials such as: pattern transformations in porous materials during the processing \citep{Singamaneni2009a,Singamaneni2009b,Singamaneni2010}; 
study of recoverable pattern transformations in cellular membranes made of shape memory polymers \citep{Li2012}; 
analysis of different patterns induced by swelling in cellular hydrogel membranes \citep{Wu2014};
and microfabrication of sub-millimetre metamaterials \citep{Dong2018}.
These examples have in comon that they do not study the kinematics nor size effects of cellular metamaterials in detail.
\cite{genovese2017} studied miniaturized additively manufactured porous biomaterials by analysing strain maps obtained from Digital Image correlation (DIC).
Other studies utilized DIC for experimental analysis of metamaterials \citep{Slann2015,Tang2015,Xu2016,carta2016,ganzosch2018,hedayati2018,Easey2019,dell2019a,dell2019b,Francesconi2020}, which again focused only on large-scale specimens.

There are also many studies in the literature that focused on other aspects of cellular metamaterials such as applications \citep{Krishnan2009,Florijn2014,Francesconi2019}, 3D printed metamaterials \citep{Mirzaali2018,Bodaghi2017,Wang2015b}, 3D cellular structures \citep{Javid2016,Yuan2017,Hedayati2019} and effects of microstructure geometry \citep{Bertoldi2008,Overvelde2012,Hu2013,Overvelde2014,Mizzi2018}, etc.



In order to study size effects in miniaturized cellular elastomeric metamaterials, and to learn whether size effects observed on large-scale specimens and in numerical simulations occur also at small scales, different specimens with millimetre sized holes and various numbers of unit cells in loading direction were processed using custom-made moulds.
\textit{In-situ} micro compression tests with optical microscopy in combination with DIC are employed to measure high resolution full-field displacement maps of the specimens with different scale ratios.
Such a detailed kinematic assessment is essential in identifying the local boundary layers, and to enable accurate validation of the size effects in the global behaviour of such metamaterials.
The novelty of this contribution is therefore in combining 
\begin{enumerate*}
\item[(1)] a study of size effects on 
\item[(2)] miniaturized cellular elastomeric metamaterials with 
\item[(3)] high kinematic accuracy using DIC.
\end{enumerate*} 
This combined analysis provides a good understanding and predictive capability of the mechanical behaviour of small scale cellular elastomeric metamaterials and a useful tool for better design for dedicated applications.
}

The current experimental study was inspired by the numerical investigation of \citet{Ameen2018}, in the experiments we, however, had to deviate from the problem studied numerically on two aspects.
The numerical specimens were infinite in the transverse direction due to the use of periodic boundary conditions.
And the boundary layer analysis was done on a homogenized solution obtained by ensemble averaging of many realizations of the microstructure inside the specimen. 
These conditions are not feasible experimentally.
\rev{Thus, new Finite Element (FE) simulations, replicating the experiments, are conducted with which the experimental findings can be objectively compared.
To this end, bulk material parameters are identified by fitting a hyper-elastic Ogden model to  tensile tests done on bulk samples of the base material.
A good match is achieved between the experimental and Direct Numerical Simulation (DNS) results in both global and local behaviour of the cellular elastomeric metamaterial specimens.}
The trend in the global stress is of the same order of magnitude as the one reported by \citep{Ameen2018}.
\rev{It is shown that the size and emergence of boundary layers strongly depend on the lateral confinement of the metamaterial, which has direct consequences for its optimal design.}
These findings indicate the influence of the boundary conditions relevant to real applications of cellular elastomeric metamaterials on their mechanical response beyond the pattern transformation, which was not addressed in idealized numerical studies in the literature. 


%
%
\section{Methodology}
\label{sec4:method}

\subsection{Specimen preparation}
\label{sec4:specimen}

\begin{figure}
\centering
\subfloat[Specimen design]{\includegraphics[height=.3\textwidth]{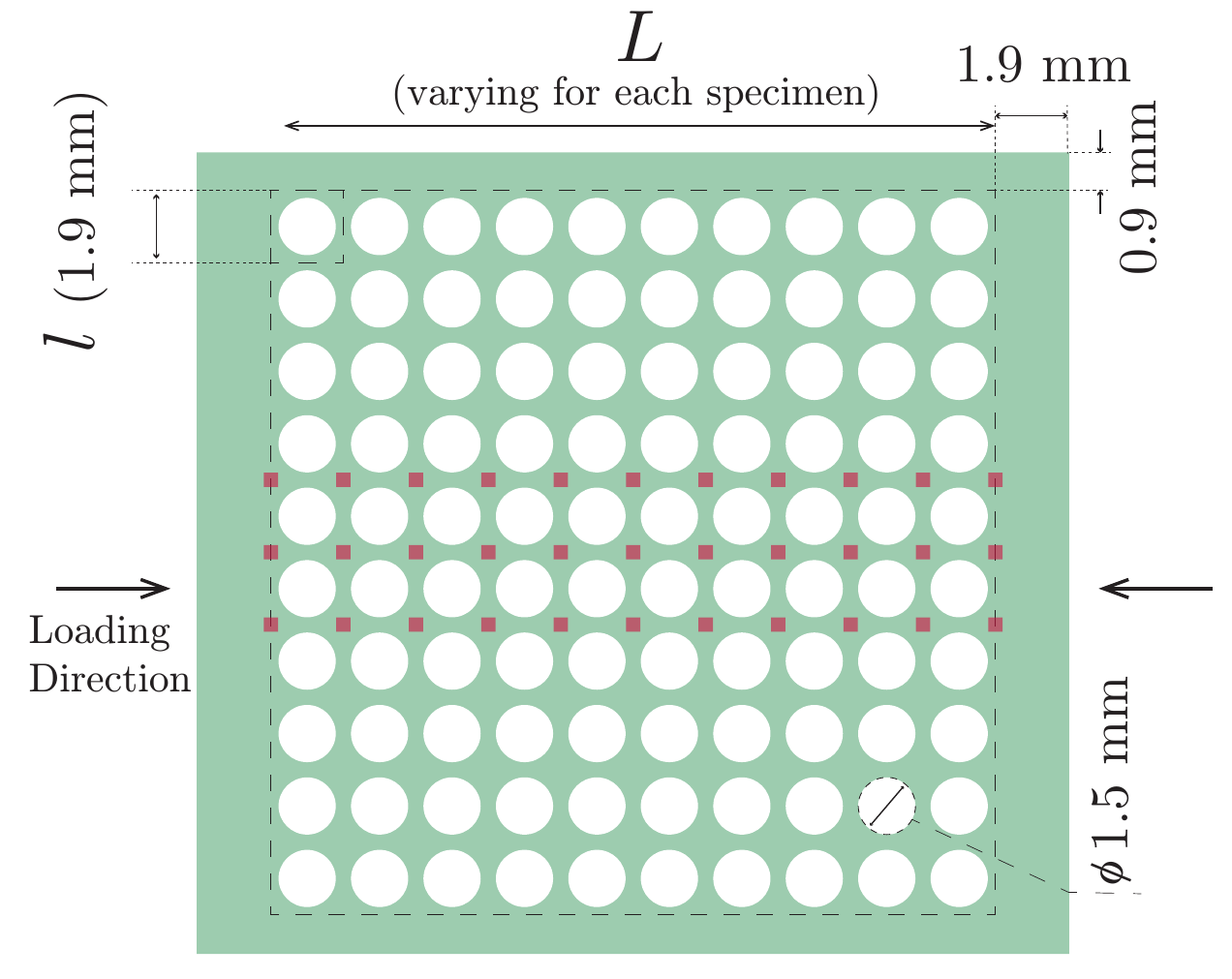}\label{fig4:sample_sketch}}\hspace{1.0em}
\subfloat[Mould]{\includegraphics[height=.3\textwidth]{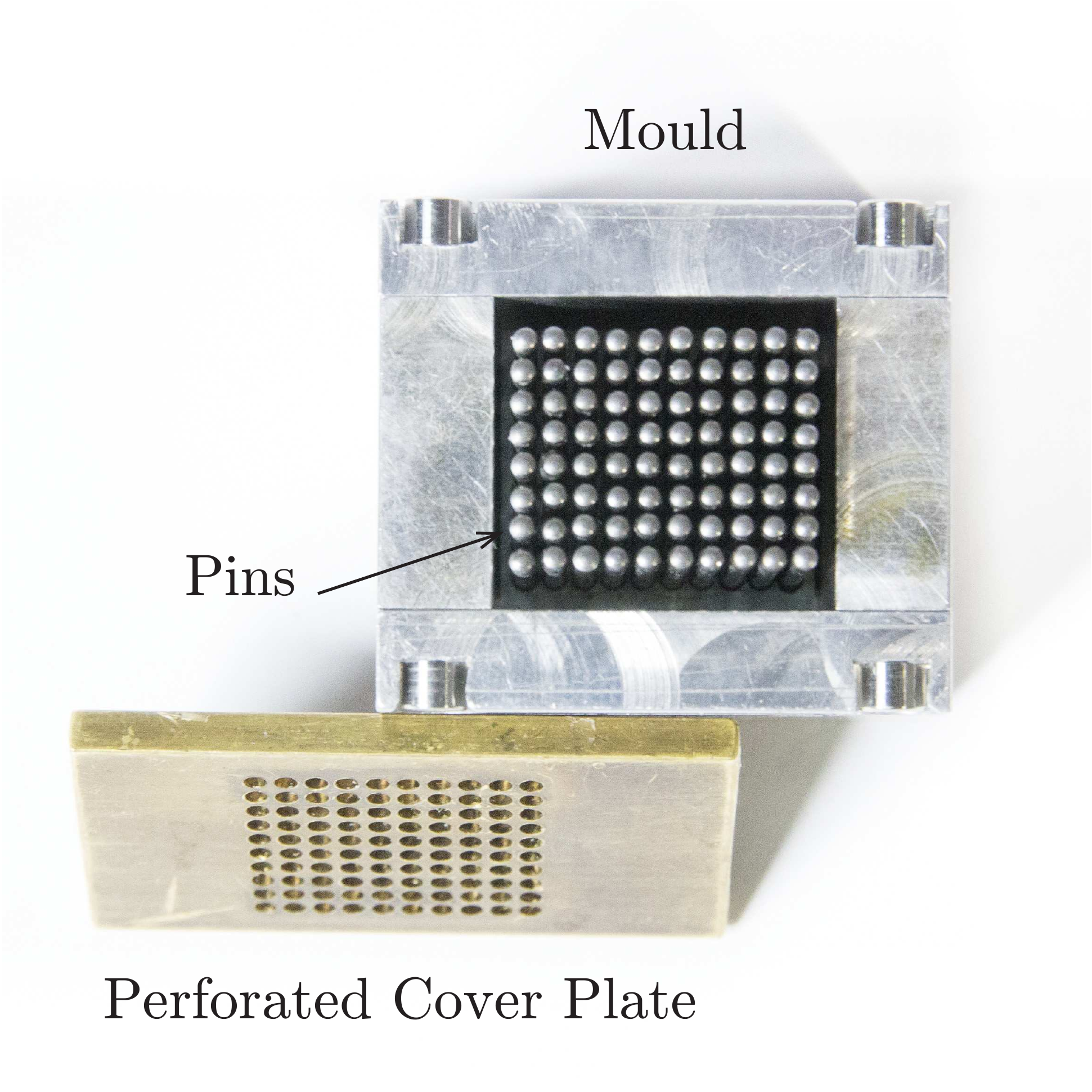}\label{fig4:mould}}\\
\subfloat[Speckle Pattern]{\includegraphics[height=.3\textwidth]{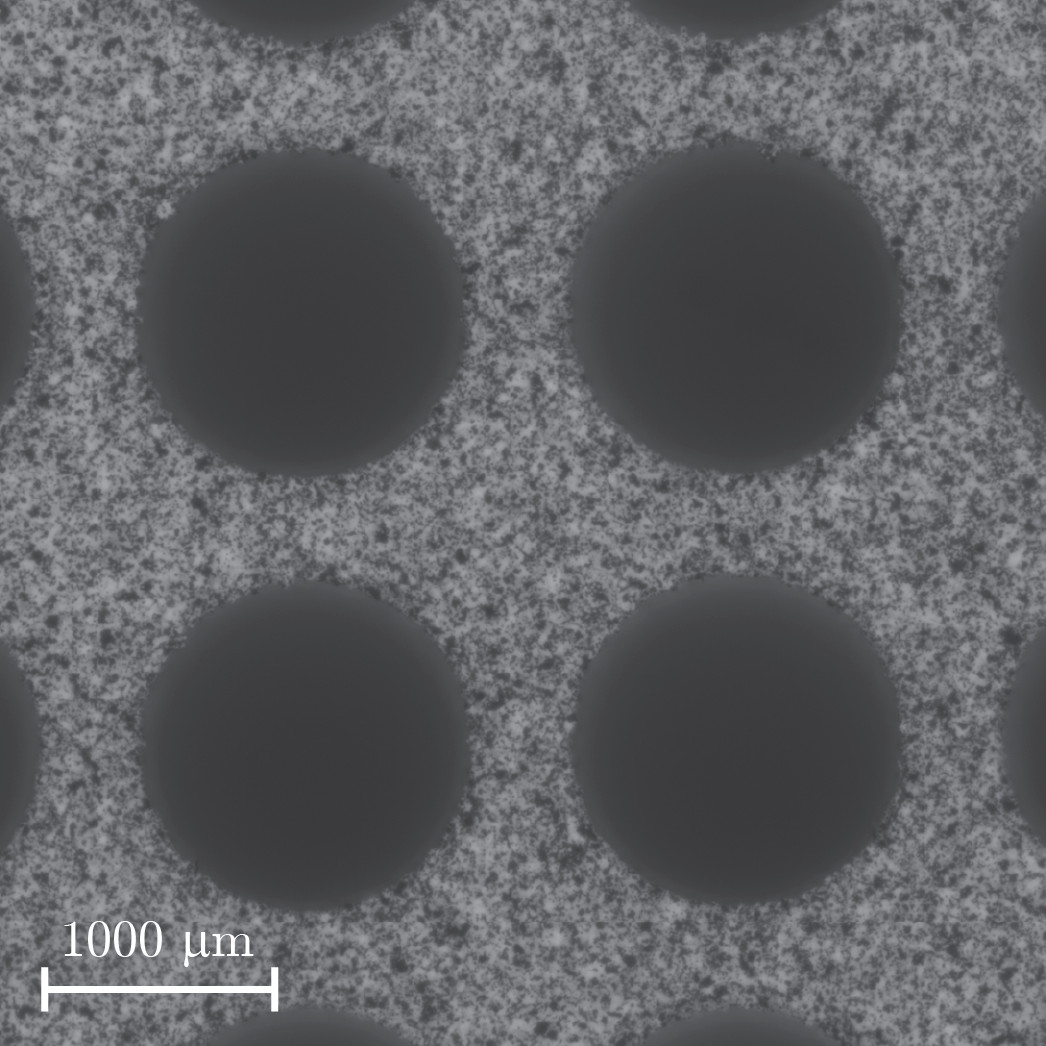}\label{fig4:speckle}}\hspace{1.0em}
\subfloat[Test setup]{\includegraphics[height=.3\textwidth]{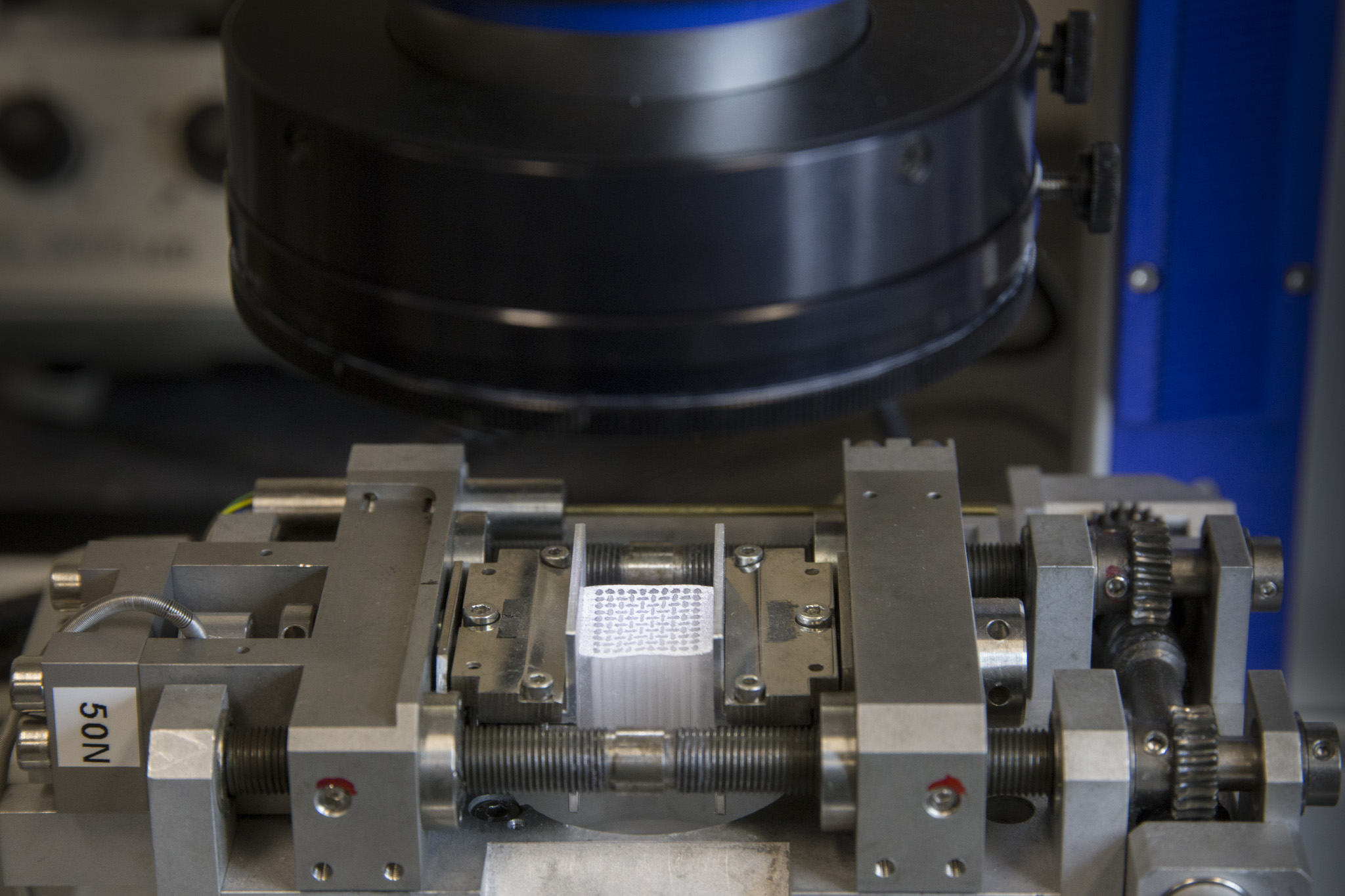}\label{fig4:test}}
\caption{Experimental set-up for \textit{in-situ} testing. \protect\subref{fig4:sample_sketch} Schematic of a $10\times10$ holes specimen (scale ratio of 10) indicating the total length of the specimen excluding the bulk side edges ($L$: the large dashed square), the size of the unit cells ($l=1.9$~mm: small dashed square), the diameter of the holes ($1.5$~mm), the size of the bulk edges and the loading direction. The red squares of $20\times20$ pixels depict the position where the local rotation of the islands (cross-shaped parts in the center of each four holes) are probed for quantifying the boundary layer size. \protect\subref{fig4:mould} The aluminium mould, with inserted pins, used for making the $10\times8$ holes specimens, along with the brass perforated cover plate. \protect\subref{fig4:speckle} Zoomed view of the speckle pattern applied on the specimens for DIC. \protect\subref{fig4:test} Test setup showing the micro compression stage, with $50~N$ load cell, underneath the microscope's objective with LED ring light mounted.}
\label{fig4:exp}
\end{figure}

Thick cellular elastomeric metamaterial specimens with millimetric circular holes have been processed with customised moulds with a rectangular stacking and micrometric spacings.
It is important to make thick specimens, and small spacings with respect to the hole diameter to induce local buckling, rather than global buckling, during the \textit{in-situ} micro-compression tests.
Note that no out-of-plane constraints can be applied to the surface of the specimens due to the DIC speckle pattern applied on the specimen surface.
All specimens have $1.5$~mm hole diameter, $1.9$~mm center to centre spacing and are $22.5$~mm thick, cf.\ Fig.~\ref{fig4:sample_sketch}.
Bulk edges of $1.9$~mm and $0.9$~mm are present in the axial and transverse directions, respectively.
The size of each unit cell is denoted $l$, whereas the length of the specimen (excluding the bulk edges) is denoted $L$, see Fig.~\ref{fig4:sample_sketch}.
The scale ratio of each specimen, $L/l$, is equal to the number of holes in the loading direction.
Specimens with scale ratios of 4, 6, 8, 10 and 12, all with 10 holes in the transverse direction, have been manufactured.
Two specimens for the scale ratios 8, 10 and 12, and a single specimen for the scale ratios 4 and 6 were made.
Based on the study of \cite{Ameen2018}, the scale ratio of 12 should be sufficient to allow for a completely developed boundary layer.
Aluminium cubic moulds with perforated  bottom surface, removable steel pins, and perforated brass cover plates are utilized, cf.\ Fig.~\ref{fig4:mould}.
Sylgard 184 with 1:10 mixing ratio was degassed for 30 minutes, then slowly poured into the moulds with inserted pins and degassed for 2 hours.
Finally, the perforated cover plates are placed to close the mould from the top and the specimens are cured for 15 hours at 70\textcelsius.
\rev{Degassing of the filled moulds should be very slow due to the small area between the pins and the high surface tension of the Polydimethylsiloxane (PDMS) mixture.}
Thus the bulk material edges are included in the specimen design to provide an outlet for the air bubbles and achieve complete degassing before the mixture viscosity increases too much during the curing process.
Note that the long curing time at an elevated temperature is necessary for thick specimens to ensure proper and homogeneous curing.
Each specimen is finalized by removing all pins, and then the cover plate.

A speckle pattern was applied on the specimens to generate a contrast as required for DIC.
Since Sylgard 184 is translucent and light reflecting, the speckle pattern consists of two layers; first a white powder (HELLING Standard-Check developer Medium Nr.~3) was sprayed to make a fine-grained background, then black India ink was sprayed using an air brush. 
Spraying at an inclined angle through a short tube, to create a spiral flow, filters out the big particles, resulting in speckles measuring roughly 30 to $80\ \mu m$, corresponding to 3 to 8 pixels as shown in Fig.~\ref{fig4:speckle}.

\subsection{\textit{In-situ} micro-compression tests}
\label{sec4:tests}

Micro-compression tests are conducted on the patterned specimens using a Kammrath \& Weiss micro tensile/compression stage with a $50\ N$ load cell.
Two aluminium T-shaped clamps are used to apply the compression uniformly in the axial direction, cf.\ Fig.~\ref{fig4:test}. 
A Zeiss Discovery.V20 stereo microscope with a PlanApo S~$0.63 \times$ objective and an Axiocam 506 mono camera with $2751 \times 2207$ pixels resolution are used to acquire images during the tests.
A $7.6 \times$ magnification is used to cover the whole area of the largest specimen.
The test is interrupted to record images at displacements corresponding to 1.4\% global strain increments up to 7\% strain, followed by 0.2\% global strain increments up to 12\% strain for all specimens.
The global strain for defining the load increments is calculated as the displacement readings of the Linear variable differential transformer (LVDT) of the compression stage, divided by the initial length of each specimen.
The test on each specimen is repeated four times. 
During the first and last test, images are acquired at $8.8 \times 10^{-5} s^{-1}$ global strain rate. The other two tests are uninterrupted and performed at $3.3\times 10^{-4} s^{-1}$ global strain rate.
Unloading during all tests is done with $3.3\times 10^{-4} s^{-1}$ global strain rate.

\subsection{Full-field displacement measurement}
\label{sec4:dic}

The images taken during the \textit{in-situ} tests are used for local DIC to extract the kinematic fields.
Images are typically affected by optical distortions of the objective lens. 
These spatial distortions are measured using the method proposed by \cite{Maraghechi2018,Maraghechi2019}.
The measured distortion field is used to correct the images taken during the \textit{in-situ} tests, which is necessary because the magnitude of the distortions is roughly 10\% of the actual displacements in the metamaterial specimens.
The kinematic fields are then determined using the VIC-2D local DIC package with 19 pixel subset and 1 pixel step size. 
The displacement fields are used to evaluate local rotation fields using the module available in VIC-2D and a smoothing filter size of 51 pixels.

\subsection{Numerical Methodology}
\label{sec4:num}

\rev{
A plane strain finite element model is made for each specimen size, using an in-house nonlinear finite element code developed in the MATLAB\textsuperscript{\tiny\textregistered} environment. 
The code employs the Total Lagrangian formulation \citep{deborst2012}, and, using the gmsh mesh generator \citep{geuzaine2009}, discretizes the specimen domain with quadratic triangular elements having three Gauss integration points. 
A standard Newton algorithm with a direct LDL\textsuperscript{T} solver for the solution of the resulting system of linear equations is used to equilibrate internal forces at each load increment. 
Since no initial geometric nor material imperfections have been adopted, after each time increment all diagonal elements of the diagonal matrix D of the decomposed current iterative stiffness matrix are checked to indicate an unstable equilibrium configuration \citep[cf.~ Section~7.1.2][]{Wriggers}. 
When a non-positive entry is encountered, the eigenvector corresponding to the lowest eigenvalue of the iterative stiffness matrix is computed and used as a perturbation of the Newton solver at a given load increment. 
The magnitude of the perturbation is increased from initially a very small value until a stable configuration is reached. 
If the perturbation was successful, the algorithm proceeds with a stable buckled configuration, otherwise the load increment is halved and the entire procedure is repeated. 
Dirichlet boundary conditions are applied on the two vertical edges perpendicular to the loading direction, whereas the two remaining edges are left free. 
PDMS is modeled with a hyper-elastic Ogden material model. 
Considering the range of different material properties reported in the literature \citep{Liu2009,Kim2011,Gerratt2013,Johnston2014}, tensile tests are performed to characterize the bulk mechanical properties of the PDMS used. The Ogden model material parameters are identified from these tensile tests as $c_1 = 0.0892$~MPa, $c_2 = 1.2537$~MPa, $m_1 = 10.0959$, $m_2 = 0.0036$ and $\kappa = 29.6716$~MPa. More details on the material characterization are given in \ref{sec4:app}.
}

%
%
\section{Results and Discussion}
\label{sec4:results}

\begin{figure}
\captionsetup[subfigure]{labelformat=empty}
\centering
\subfloat[$4\times10$]{\includegraphics[trim=210 0 210 0,clip,height=.4\textwidth]{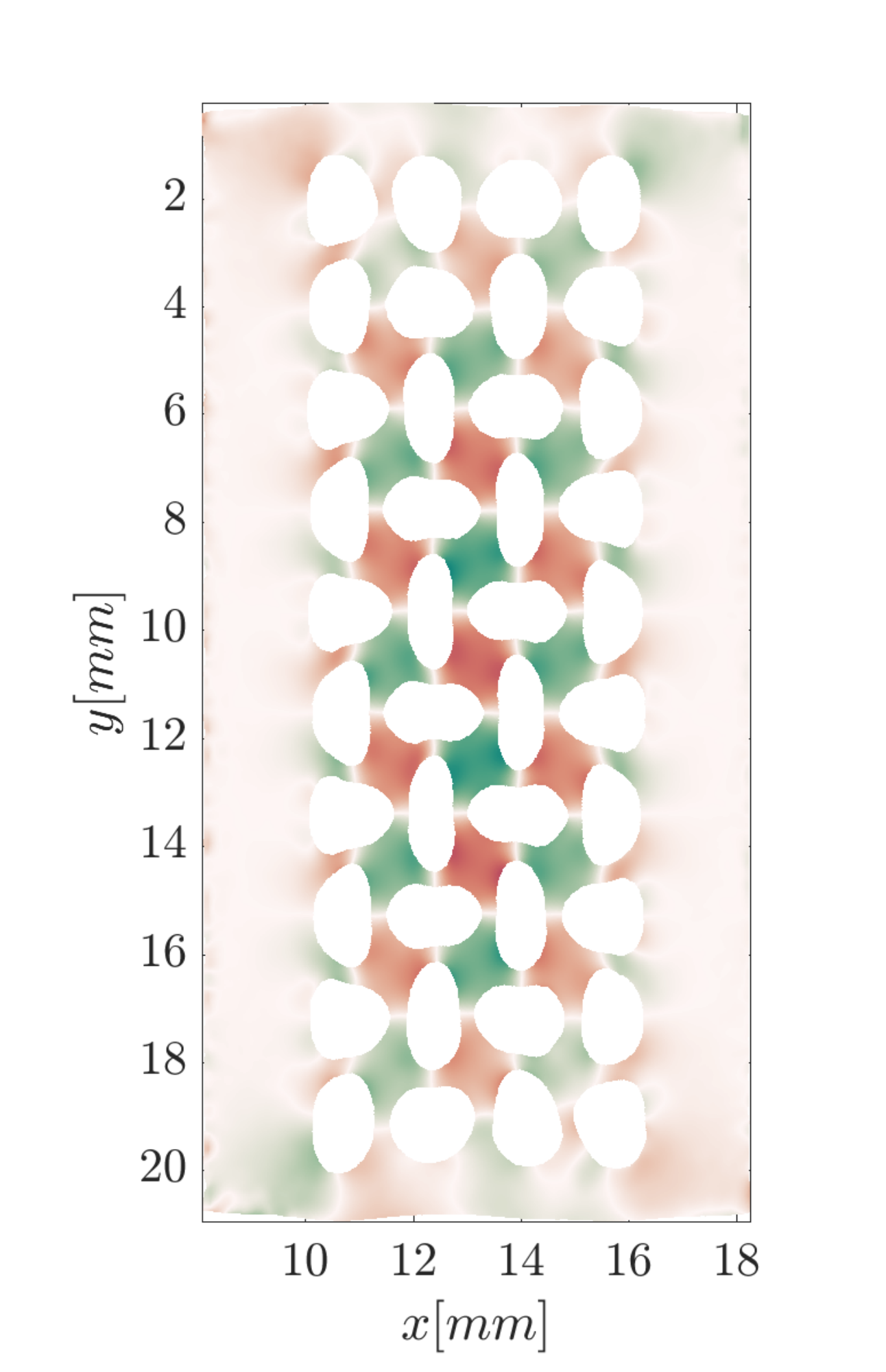}}
\subfloat[$6\times10$]{\includegraphics[trim=180 0 180 0,clip,height=.4\textwidth]{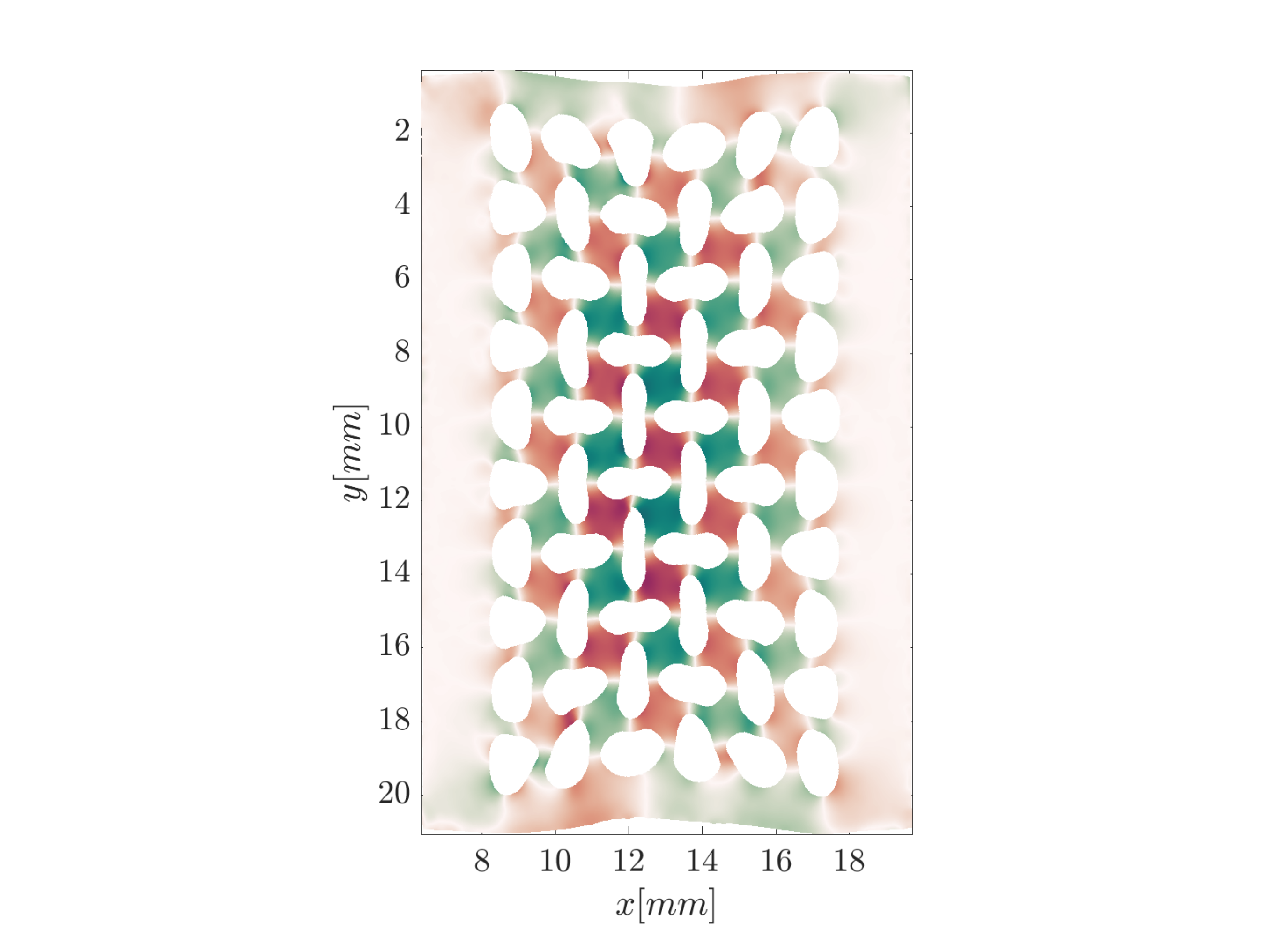}}
\subfloat[$8\times10$]{\includegraphics[trim=140 0 140 0,clip,height=.4\textwidth]{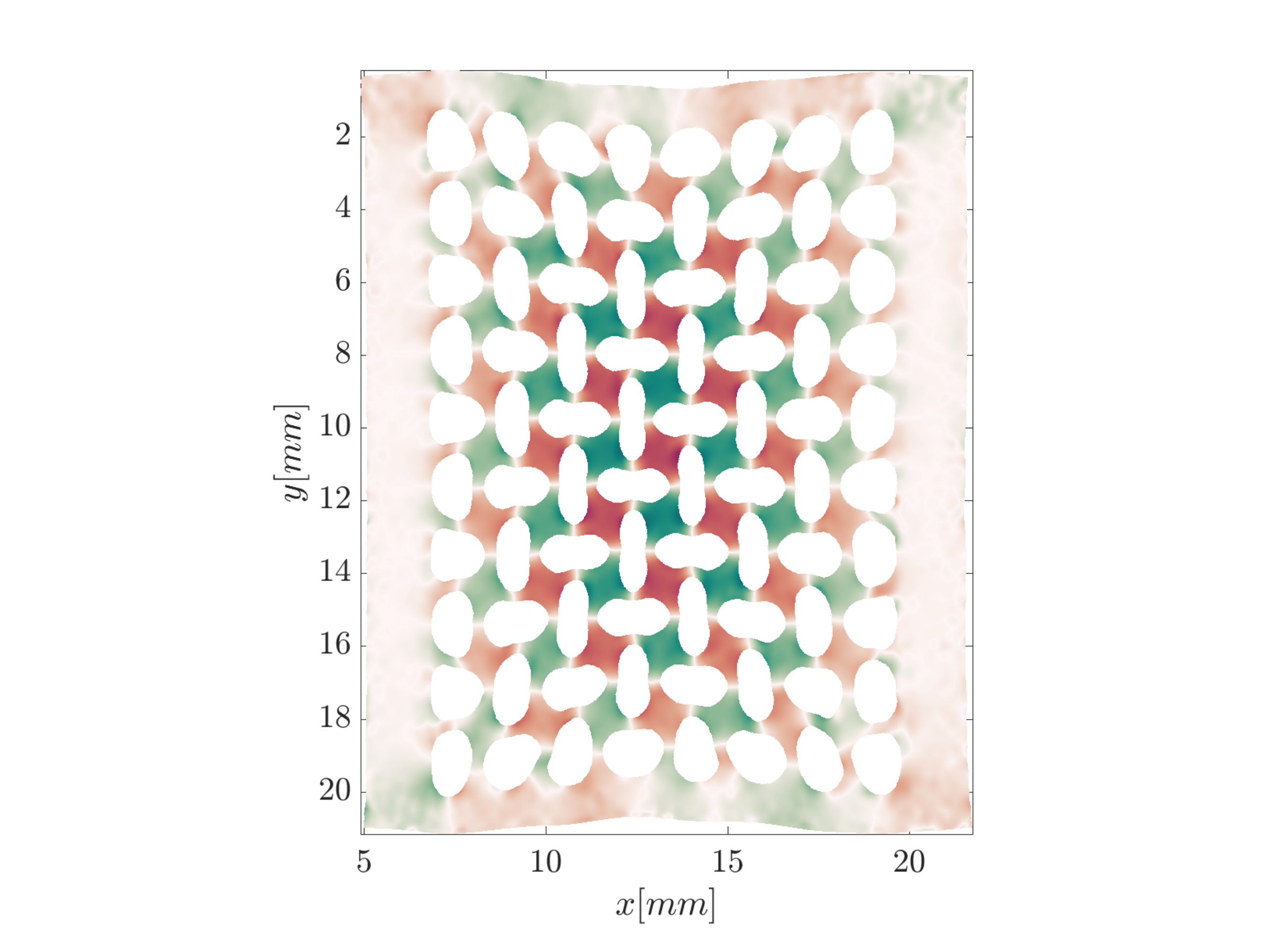}}\\
\subfloat[$10\times10$]{\includegraphics[trim=100 0 100 0,clip,height=.4\textwidth]{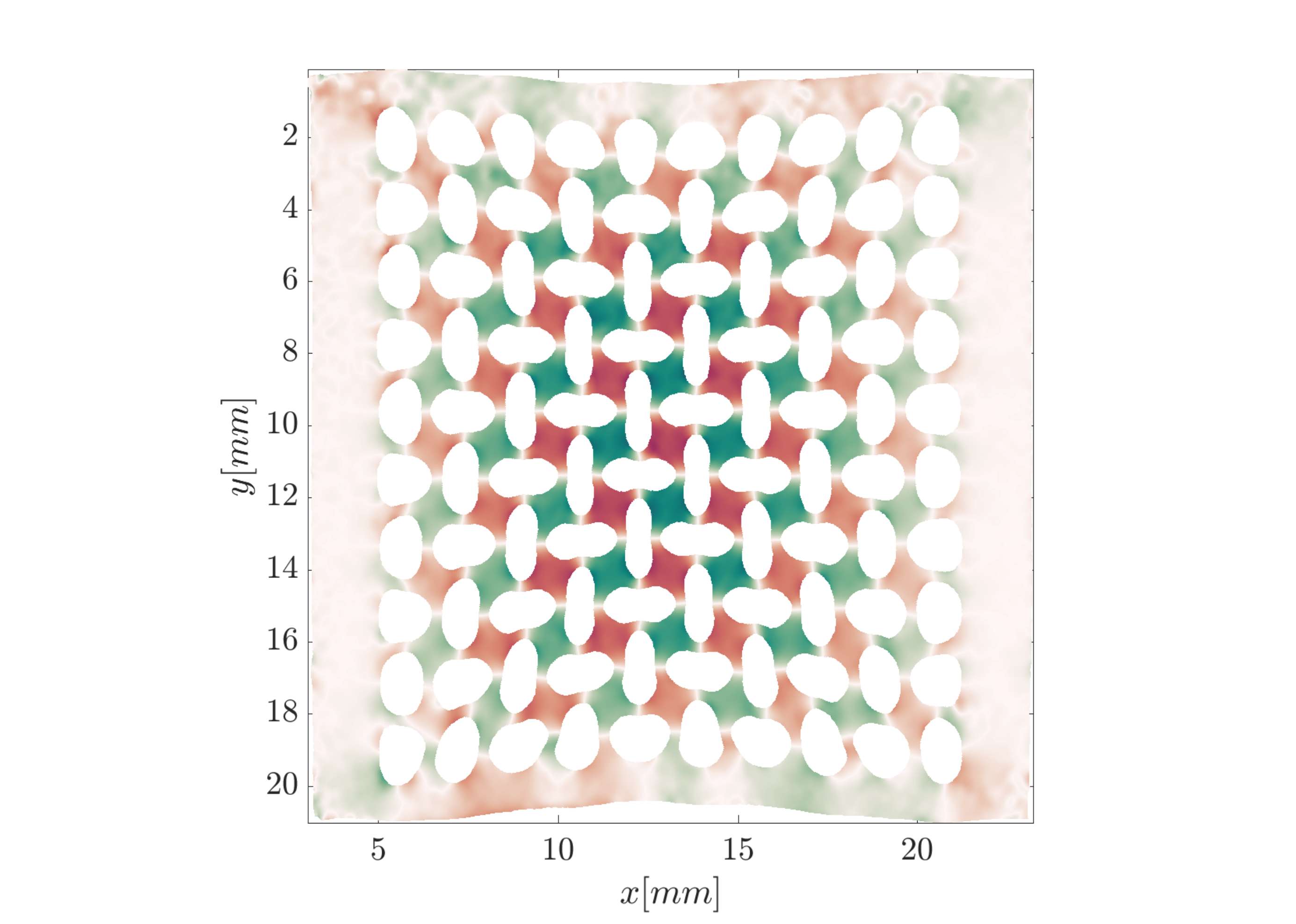}}
\subfloat[$12\times10$]{\includegraphics[trim=60 0 60 0,clip,height=.4\textwidth]{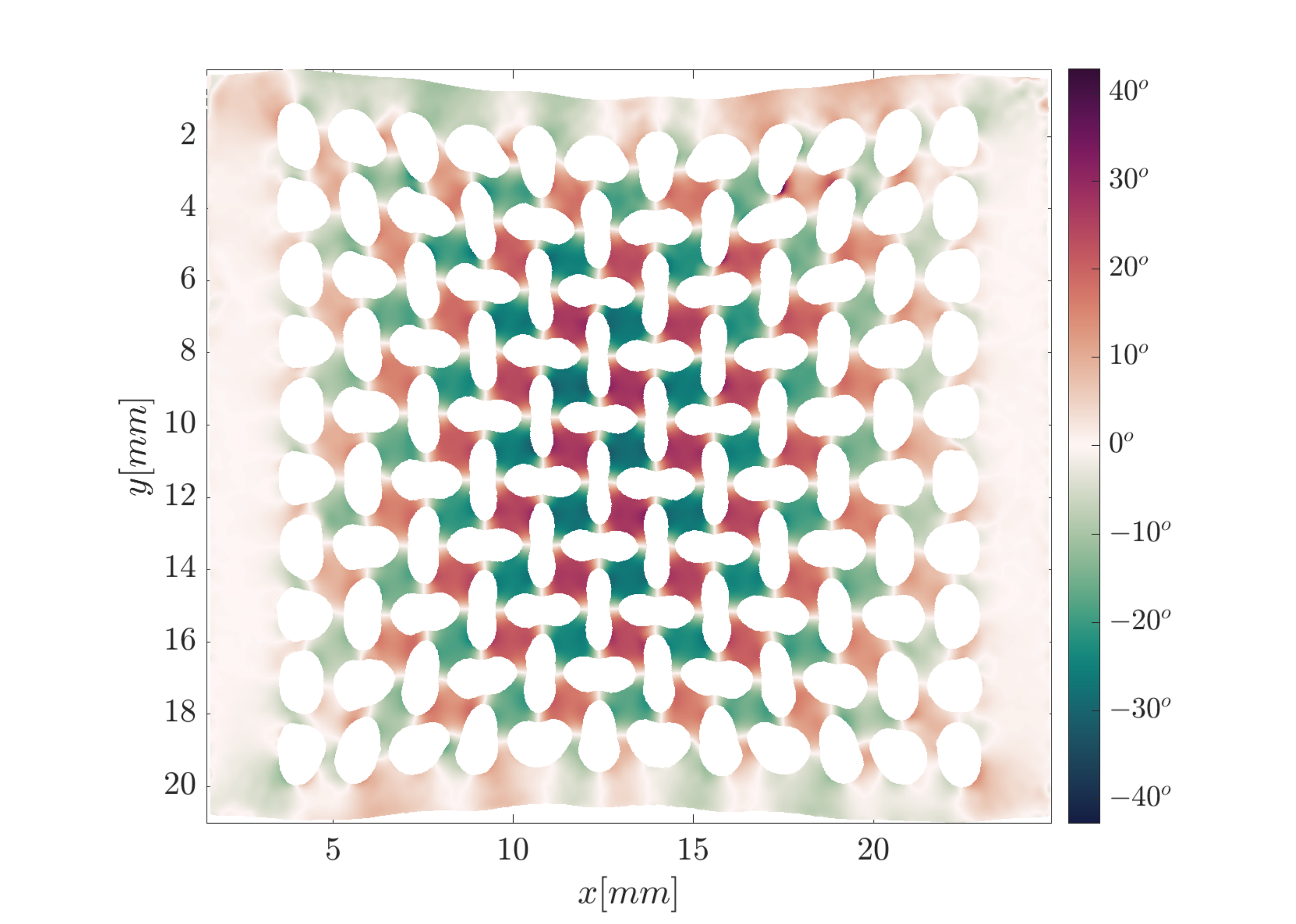}}
\caption{Local rotation fields inside a specimen for each scale ratio, at 1.3\% global strain after local buckling, plotted on the deformed configuration. The displacement fields used for calculating the deformations and the local rotations are obtained by local digital image correlation.}
\label{fig4:rot_field}
\end{figure}

Local rotation fields are obtained from displacement fields obtained by local DIC, as described in Section \ref{sec4:dic}, and are shown in Fig.~\ref{fig4:rot_field} in the deformed configuration for one specimen of each scale ratio at 1.3\% global strain after  buckling.

\subsection{Global Response}
\label{sec4:global}

In order to accurately measure the applied displacement on the specimens, and thus the nominal strains, the DIC data are used.
In order to properly quantify the displacement to achieve the same global strain for specimens of different scale ratios, the exact geometry of the specimens has to be accounted for.
To this end, the global displacement is determined at the outside of the first and the last unit cells (i.e.\ excluding the bulk edges), where the applied displacement on each side is averaged over the width of the specimen.
The global nominal strain is then calculated as the ratio of the relative displacement between the two sides and the initial distance, $L$.
The global nominal stress is computed as the ratio of the global compressive force and the initial cross sectional area of each specimen. 
This approach is used for both experimental and numerical analyses.

\begin{figure}
\centering
\subfloat[Stress--strain curves]{\includegraphics[width=.75\textwidth]{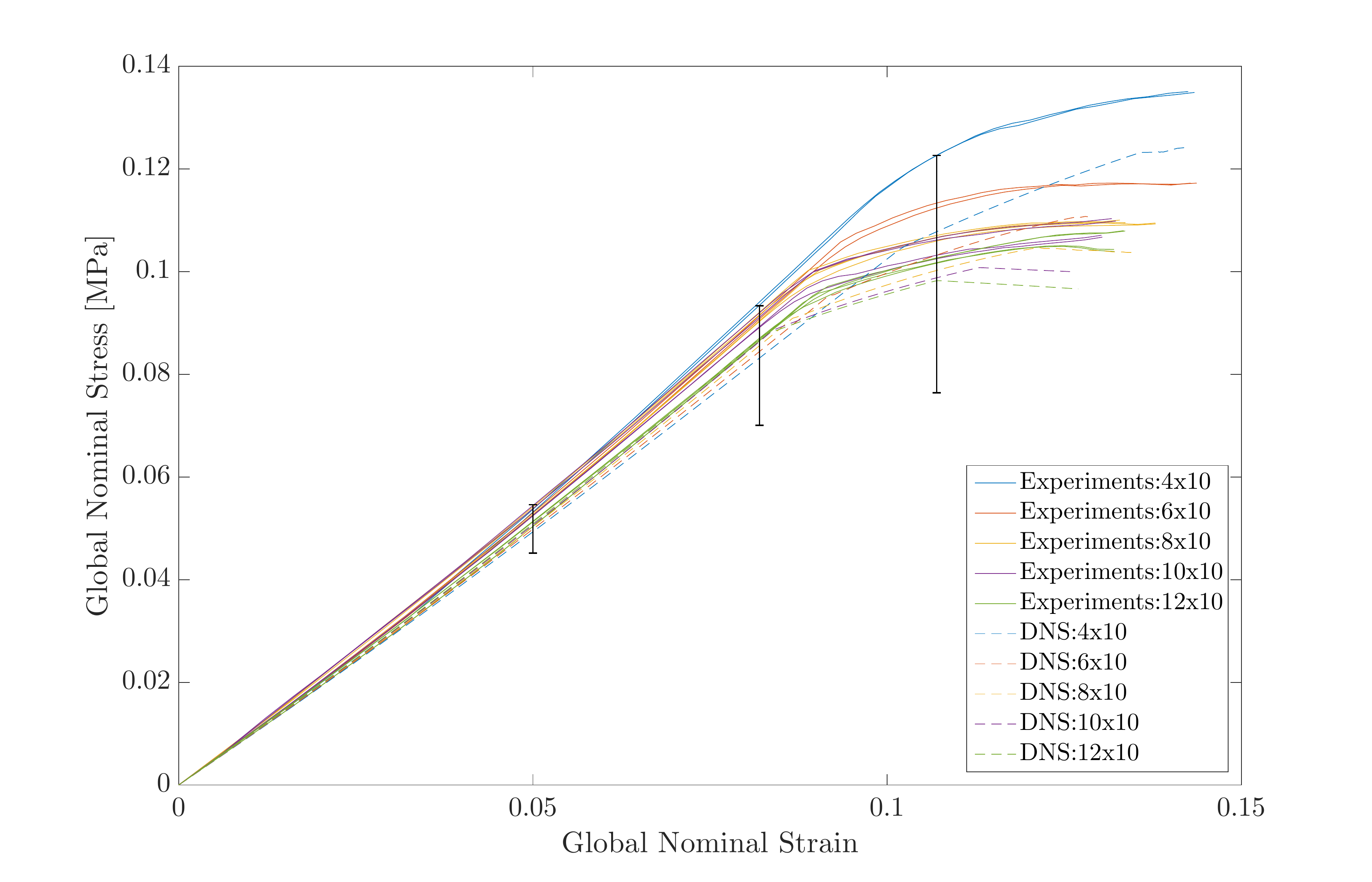}\label{fig4:fd}}\\
\subfloat[Buckling strain]{\includegraphics[width=.45\textwidth]{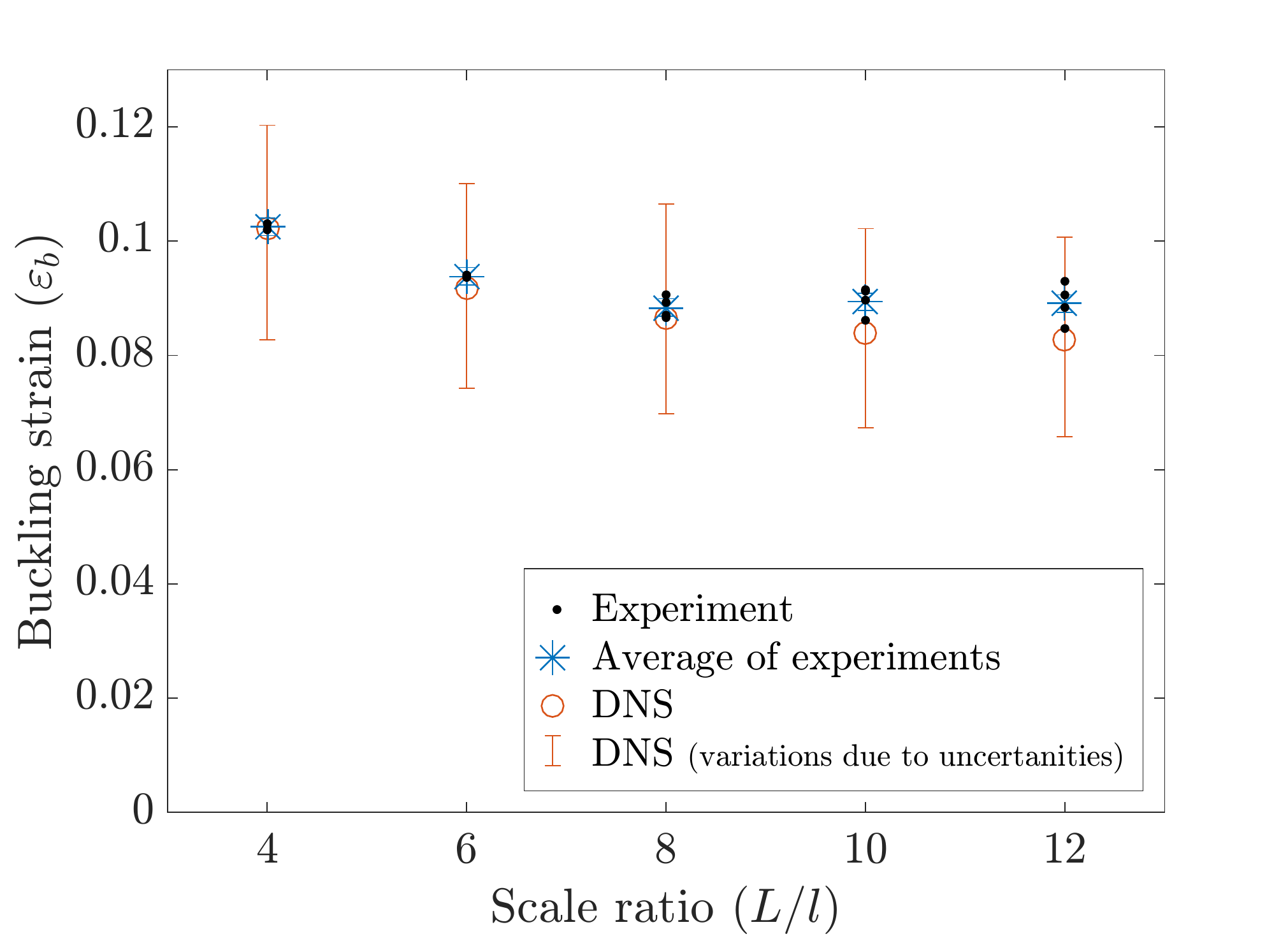}\label{fig4:buck_strain}}\hspace{1.0em}
\subfloat[Post-Buckling stress]{\includegraphics[width=.45\textwidth]{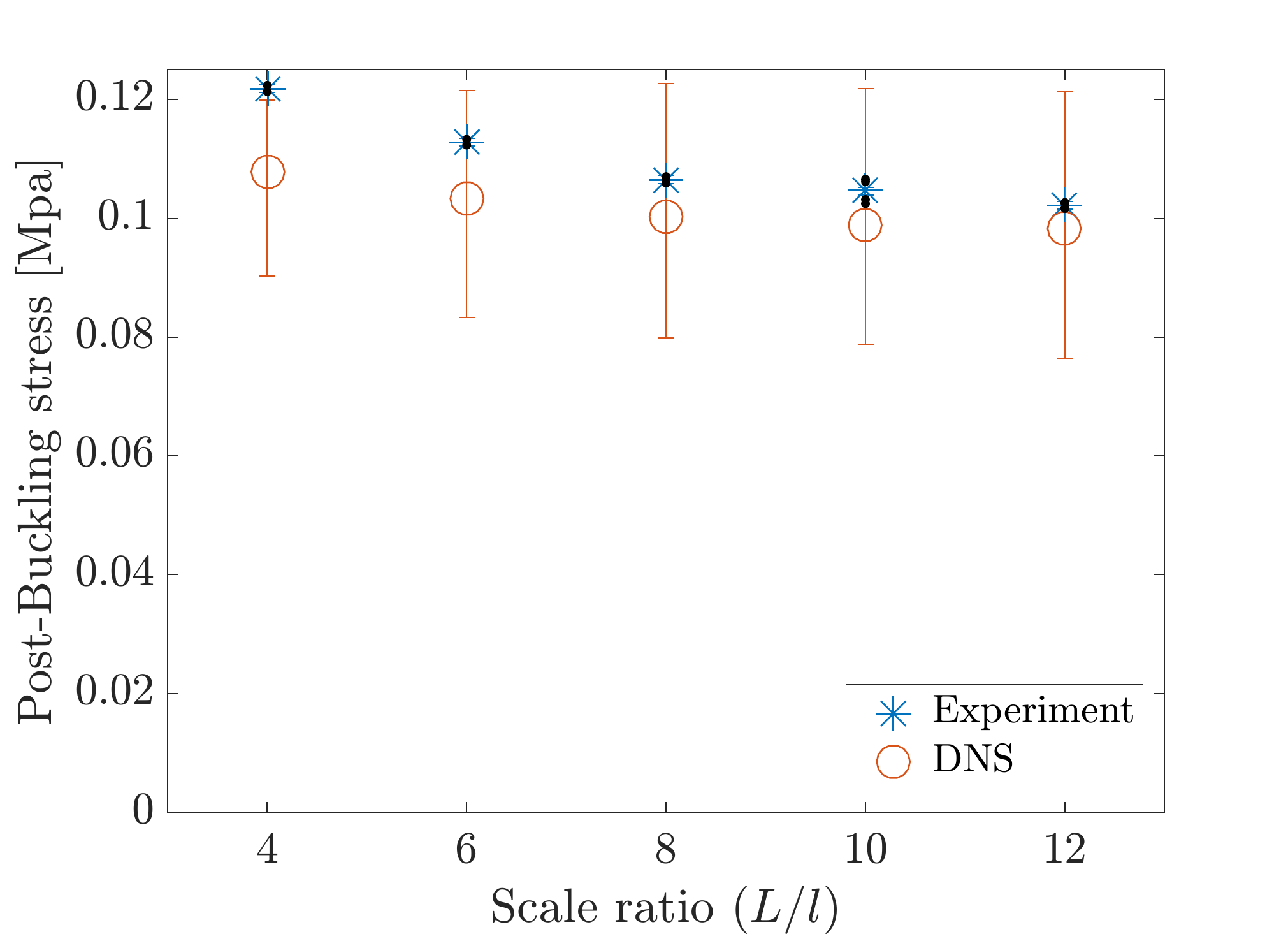}\label{fig4:post_buck}}
\caption{\rev{Global response of specimens of different scale ratios, comparing experimental and Direct Numerical Simulation~(DNS) results. \protect\subref{fig4:fd} Global nominal stress--strain curve. Results of all tests for each scale ratio are depicted in the same color and the error bars reflect the uncertainties in the simulations, resulting from the geometrical variations. \protect\subref{fig4:buck_strain} Buckling strains, calculated as the strain at which the maximum curvature occurs in the stress--strain curves. \protect\subref{fig4:post_buck} Post-buckling stresses determined at a global strain of 10.7\% (maximum strain before global buckling of the largest specimen). The error bars on the experimental data in \protect\subref{fig4:buck_strain} and \protect\subref{fig4:post_buck} reflect the variability in all tests done, while the error bars on numerical data reflect the uncertainties in the simulations, resulting from the geometrical variations. The small black dots in \protect\subref{fig4:buck_strain} and \protect\subref{fig4:post_buck} represent the results of each individual test, while the blue stars the average for each scale ratio.}}
\label{fig4:global}
\end{figure}

Fig.~\ref{fig4:fd} depicts the global stress-strain curves of different specimens for both the experiments and the numerical simulations.
The results of all tests for each scale ratio are depicted in the same color for the sake of readability.
Optical inspection of the specimens under the microscope showed variations of approximately 3~\% in hole diameter.
Therefore extra simulations were conducted considering $\pm3~\%$ variation in the diameter of the holes.
\rev{Variations of the hole diameters are taken as an example representing different parameters contributing to the uncertainties in the experimental results, such as positional tolerance of the holes, material and loading imperfections, etc.}
The error bars at three different strain levels in Fig.\ \ref{fig4:fd} depict the range of stresses observed in these simulations.
Note that the somewhat higher stiffness observed in the experiments is in line with the optical inspection of the specimens pointing to smaller holes than the design value.
The pre-buckling stiffness values of different specimens agree well, as expected for cellular metamaterials, for which the pre-buckling stiffness should be independent of the scale ratio.
There is, however, a small systematic variation in the initial stiffness due to the bulk edges on the transverse sides, which was confirmed in numerical simulations with and without these side bulk edges.

\rev{The onset of the local buckling, resulting in the pattern transformation, is found based on the maximum curvature of the stress-strain curves of Fig.~\ref{fig4:fd} for both the experimental and numerical results.
Fig.~\ref{fig4:buck_strain} shows the global strain at which microstructural buckling occurs in each specimen, which is denoted by $\varepsilon_b$.}
The average buckling strain for all tested specimens of a certain scale ratio are depicted in this curve, and the error bars are attained as follows.
The differences between the buckling strains for all 16 tests and the average of the corresponding scale ratio are calculated.
The standard deviation of these disparities gives a global measure of the uncertainty in the buckling strain measurements, and is therefore used as the error bar.
Based on the observed geometrical variations, the error bars on the numerical results are based on two sets of simulations considering $\pm3~\%$ variation of the hole diameters.
The buckling strains match well between the experimental and numerical results (Fig.~\ref{fig4:buck_strain}).
Note that a small variation in the geometry of the specimens results in a large shift in buckling strains. 
The asymptotically decreasing trend, however, remains uninfluenced.
\rev{The maximum relative error is 7.2\% while the minimum relative error range indicated by the error bars of the DNS results is 19.0\%.}

Fig.~\ref{fig4:post_buck} depicts the post-buckling stress against the scale ratio, taken from both the experimental and numerical results.
\rev{The stress is determined at 10.7\% of applied global strain for all specimens.
This is the maximum strain at which all specimens exhibit the local buckling pattern while none of them has yet gone through global buckling, i.e.\ the maximum strain prior to the global buckling of the largest specimen with scale ratio 12. }
The error bars are obtained as in Fig.~\ref{fig4:buck_strain}. 
A nearly constant trend is observed in the upper bound of the error bars due to the fact that the simulation of the two smallest specimens with a $3 \%$ smaller hole diameter did not yet reveal local buckling at the imposed 10.7\% global strain.
Taking this into account, the numerical and experimental results show a similar asymptotically decreasing trend.
\rev{The maximum relative error is 11.6\% while the minimum relative error range indicated by the error bars of the DNS results is 20.0\%.}
The larger size effect in the experiments is due to the fact that the onset of global buckling in the larger specimens is more gradual, influencing the stresses at the considered strain.

\rev{The global post-buckling stress in the current study shows a 19\% decrease from scale ratio 4 to 12.
This value is ower than the 29\% observed in the idealized numerical study of \cite{Ameen2018}.
The difference is due to the relative size of the holes with respect to the unit cell as well as different boundary conditions; recall that no free boundaries nor bulk edges are present in the infinitely wide specimens considered in the work of \cite{Ameen2018}.}

\subsection{Local Behaviour}
\label{sec4:local}

The boundary conditions applied to the cellular elastomeric metamaterial specimens obviously influence the buckling-induced pattern close to the edges, which in turn result in the size effects discussed above.
\citet{Ameen2018} studied this effect by numerically identifying the boundary layer in which the normal strain in the loading direction is affected by the edge, using a homogenized solution.
This homogenized solution was calculated from an ensemble average of many realizations of the same periodic microstructure, where the microstructure was shifted relative to the specimen geometry.
Since it is impossible to use the same procedure for experimental results (for which only one realization is measured), a different approach for characterising the boundary layer thickness is proposed here. 
The local buckling of the ligaments between the holes results in rotation of the islands, leading to the patterns shown in Fig.~\ref{fig4:rot_field}.
When the specimen is loaded beyond the buckling point, a significant increase in the rotation of the islands results in a pronounced pattern.
This also applies to the spatial variation of the pattern at each global strain state, i.e.\ the rotation of the islands is larger in the specimen center, where the pattern is more developed. 

\begin{figure}
\centering
\subfloat[Experimental mean rotation angle]{\includegraphics[width=.45\textwidth]{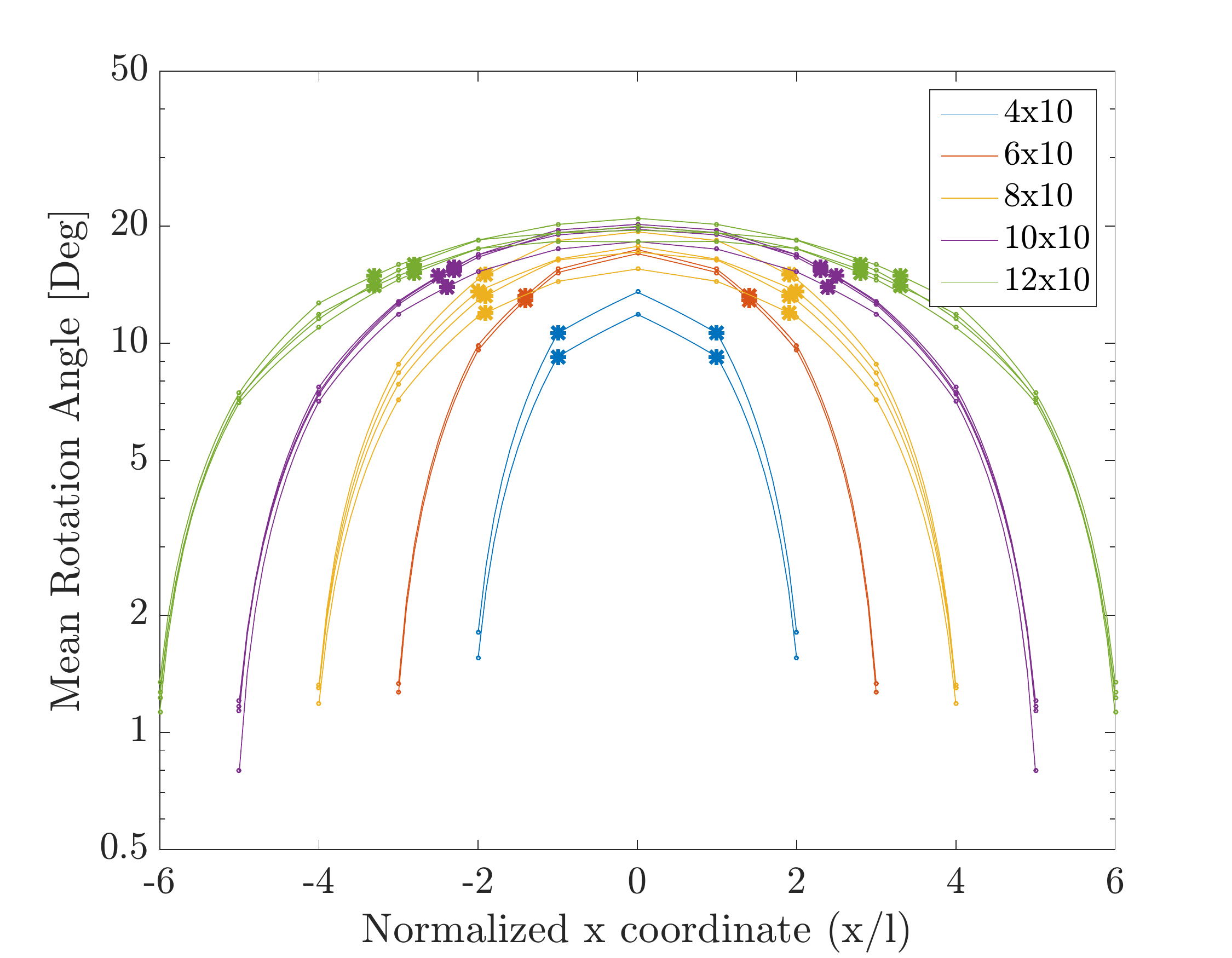}\label{fig4:rot_exp}}\hspace{1.0em}
\subfloat[DNS mean rotation angle]{\includegraphics[width=.45\textwidth]{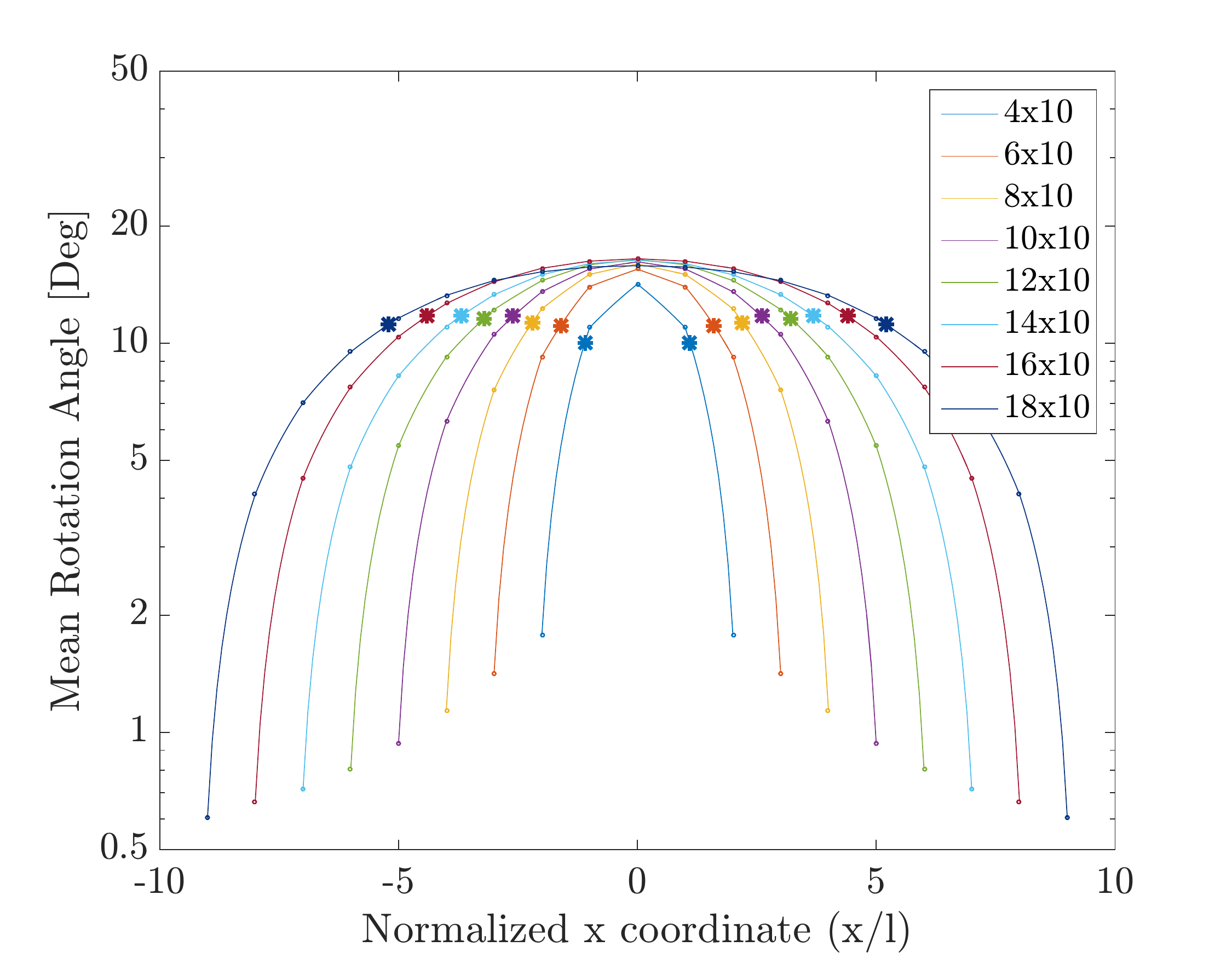}\label{fig4:rot_dns}}\\
\subfloat[Boundary layer size]{\includegraphics[width=.6\textwidth]{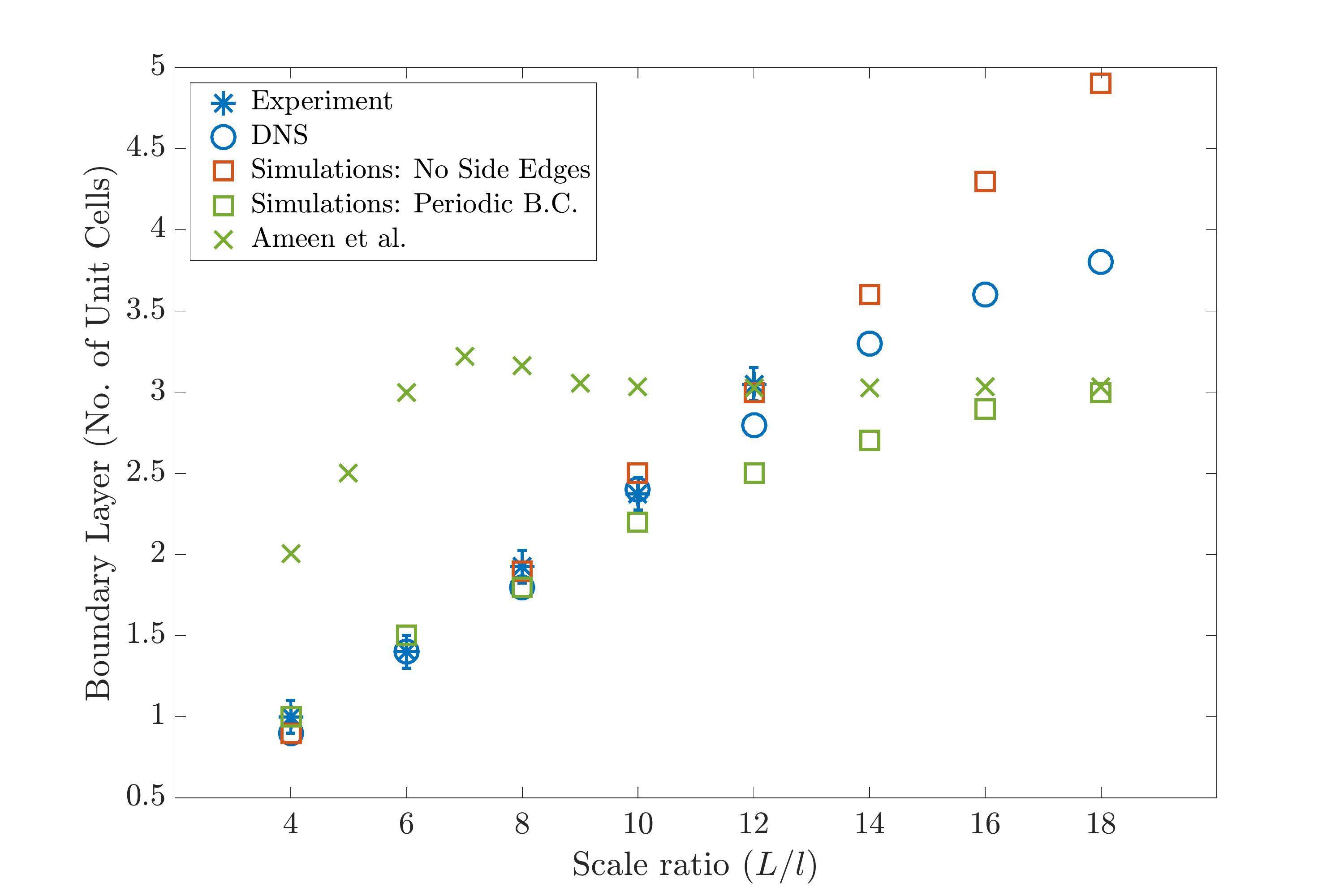}\label{fig4:BL}}
\caption{Local behaviour of specimens of different scale ratios based on experimental and Direct Numerical Simulation~(DNS) results. Rotation angles averaged over the two central unit cells against the horizontal coordinate for \protect\subref{fig4:rot_exp} experimental results, where results of all the tests for each scale ratio are depicted in the same color, and \protect\subref{fig4:rot_dns} numerical results, at a relative global strain of $\varepsilon_b+0.013$. The stars are depicting 70\% of the dynamic  range of each curve for quantification of the boundary layer size. Graphs in \protect\subref{fig4:rot_exp} and \protect\subref{fig4:rot_dns} are plotted on logarithmic vertical scales to better visualize the boundary layers.  \protect\subref{fig4:BL} Size of the boundary layer, in terms of the number of unit cells, determined from the rotation of the islands, for the experimental and numerical data as well as for the simulations by \cite{Ameen2018}. The simulations are extended to a scale ratio of 18 to reveal the converged size of the boundary layer.}
\label{fig4:rotation}
\end{figure}

Therefore, in order to quantify the size of the boundary layer, the rotation of the islands is exploited as the quantity of interest.
To avoid the influence of the free side edges, only the two central unit cell rows are taken into consideration, see Fig.\ \ref{fig4:sample_sketch}.
A unique rotation angle is determined for each island by averaging the local rotations in a window of $20 \times 20$ pixels in the center of each island. 
The little red squares in Fig.\ \ref{fig4:sample_sketch} indicate the positions where the local rotations are probed. 
The absolute values of the rotation angles of the islands are averaged over each column (i.e.\ in the transverse direction) and plotted against the row number (i.e.\ normalized axial coordinate) in Fig.~\ref{fig4:rot_exp} for the experimental and Fig.~\ref{fig4:rot_dns} for the numerical data.
Each curve is evaluated at a relative global strain increment of 0.013 after the onset of local buckling at the corresponding scale ratio ($\varepsilon_b+0.013$).
The curves in Fig.~\ref{fig4:rot_exp}~and~\subref*{fig4:rot_dns} are plotted by mirroring the rotation angles with respect to the central island and averaging the values on both sides. 
\rev{By these means the small variations in the rotation angles, due to the antisymmetry of the buckling pattern, are averaged out and smooth symmetric curves are attained.}
Values corresponding to non-integer number of holes/scale ratios are based on linear interpolation of the local rotation of islands to give a quasi-continuous measure for quantifying the boundary layer thickness. 
Each curve in Figs.~\ref{fig4:rot_exp} and \ref{fig4:rot_dns} is probed at 70\% of its dynamic range and the $x$ coordinate of each intersection defines the size of the boundary layer.
\rev{The reason behind the choice of this threshold is explained below.}
The resulting boundary layer size is plotted against the scale ratio in Fig.~\ref{fig4:BL}.
The experimental and numerical results match adequately, although the trend is not yet converged for the specimens with scale ratio 12.
Note that it is experimentally challenging to perform the tests on specimens with larger scale ratio, due to the necessity of thicker specimens to prevent global buckling making the specimen processing infeasible. 
And also due to the difficulty of attaining high spatial resolution of the displacement and thus rotation fields inside the islands, while capturing the complete surface of the specimens within the field of view.  

Three more numerical simulations are conducted to analyse larger specimens.
At a scale ratio of 18, the boundary layer size almost levels out, showing an asymptotic trend corresponding to a boundary layer size of roughly 4 unit cells.
Extending the simulations to bigger  scale ratios is infeasible since global buckling occurs almost immediately after local buckling, preventing analysis of the rotation fields before global buckling.
Note that the presented results are all determined at $\varepsilon_b+0.013$, which guarantees that all specimens up to scale ratio 18 are not globally buckled.

Results of two additional sets of numerical simulations with different lateral constraints are included in Fig.\ \ref{fig4:BL} as well.
In both cases the bulk side edges in the loading direction and 10 unit cells over the specimen width are employed.
First, specimens with different scale ratios but no bulk side edges are analysed, which imposes no lateral constraint on the metamaterial.
Second, infinitely wide specimens are computed by means of periodic boundary conditions, which applies a high level of lateral constraint on the metamaterial.
\rev{Note that these are the exact same boundary conditions as in \cite{Ameen2018}, thus providing a suitable reference for choosing a proper threshold for probing the boundary layer thickness. 
As long as the applied boundary conditions are the same, it is expected that the saturated boundary layer thickness for large scale ratios is the same, regardless of the definition and method chosen for its evaluation.
Accordingly the 70\% threshold, mentioned above, is chosen, to have the boundary layer thickness of these numerical simulations to match those of \cite{Ameen2018}, for large scale ratios.
Note, however, that they used a different measure, i.e.\ maximum curvature of the axial component of deformation gradient tensor, to assess the boundary layer size in contrast to this study.
The boundary layer size in the study of \cite{Ameen2018} is evaluated from a homogenized solution based on many shifted realizations of the microstructure, in many of which the holes cut through the edges.
For small scale ratios, this results in a different boundary layer thickness evaluated based on one single realization (no microstructural shift) and the thickness assessed from the homogenized solution.
Moreover, in their study, the boundary layer thickness is chosen to be half the specimen thickness for scale ratios upto 6, which is not translatable to the definition of boundary layer chosen in the current study.
These points explain the difference between the simulations of the current study with periodic boundary conditions and results of \cite{Ameen2018}.}

It is observed that the different levels of lateral confinement directly influence the scale ratio at which the boundary layer size converges, see Fig.\ \ref{fig4:BL}.
The experimental specimens of the current study are in between these two limit cases of lateral confinement, which explains why the boundary layer size was not fully converged in the experimental data.

The presented results demonstrate that relatively large boundary layers emerge in cellular elastomeric metamaterials, which result in considerable size effects in specimens of finite size.
These effects are tunable to a certain degree by changing the constraints put on the specimens in the transverse direction, which can be exploited for customized designs of these metamaterials in various engineering applications.

%
\section{Summary and Conclusions}
\label{sec4:conclusion}
\rev{Cellular elastomeric materials exhibit pattern transformations beyond a critical load, entailing internal buckling of their microstructure, which dramatically influences their global mechanical response.
Potential applications of cellular elastomers, such as soft robotics, typically require small specimens (millimetre sized microstructural features) and comparable macroscopic sizes, which make them sensitive to the applied boundary conditions, exhibiting considerable size effects.
To experimentally analyse the size effects in such materials, as previously observed in a numerical study of \cite{Ameen2018}, a systematic study has been conducted in this contribution.}
To remain consistent with potential applications of cellular elastomers, specimens with millimetre sized holes have been manufactured using custom-made moulds. 
The size of the specimen with respect to hole size, i.e.\ ``scale ratio'', was varied while keeping the hole size fixed.
\textit{In-situ} micro compression tests in conjunction with optical microscopy and digital image correlation have been conducted to attain full-field kinematic measurements.
Such measurements are essential to accurately assess the global stress--strain behaviour of the specimens and to evaluate the local kinematics during pattern transformation, which allows to extract the boundary layer thickness.
Unlike done in the numerical study mentioned above, real specimens are by definition finite in size and ensemble-average homogenized solutions are not applicable, making a direct comparison challenging.
To overcome this difficulty, finite element simulations of finite sized specimens have been conductedd and analysed in parallel to the experiments.
Tensile tests have been performed to characterize the bulk material, allowing to identify the Ogden hyper-elastic material parameters.
The experimentally observed geometrical variations within the specimens have been incorporated to assess the uncertainties characterising the real specimens.

The global stress--strain behaviour, local buckling strain and post-buckling stress for different scale ratios adequately agree between numerical and experimental results, taking into account the specimen variations and uncertainties.
The global stress shows a trend of the same order of magnitude as the one reported by \cite{Ameen2018}, while the difference is mainly due to the very different boundary conditions in the two studies.
In order to determine the size of the boundary layer, the spatial variation of the rotation angle of cross-islands inside each specimen was studied.
The numerical and experimental results agree well, although the boundary layer size did not yet converge to an asymptotic value for the maximum scale ratio used in the experiments. 
The numerical results indicate convergence of the boundary layer size towards approximately 4 unit cells.
Additional numerical investigations, where the lateral constraint acting on the specimens was varied, reveal that the level of lateral confinement of the cellular elastomeric metamaterials strongly influences the boundary layer and its convergence to a saturated value.
\rev{Relaxation of the lateral constraints results in larger boundary layers and larger scale ratios for which the boundary layer size has converged.}
These novel insights are instrumental for controlling the boundary layer size and size effects in designs based on cellular metamaterials.
%
%
\section*{Acknowledgements}
The research leading to these results has received funding from the European Research Council under the European Union's Seventh Framework Programme (FP7/2007-2013)/ERC grant agreement \textnumero~[339392].

\appendix
\section{Material Characterization}\label{sec4:app}

\begin{figure}
	\centering
	\includegraphics[width=.6\textwidth]{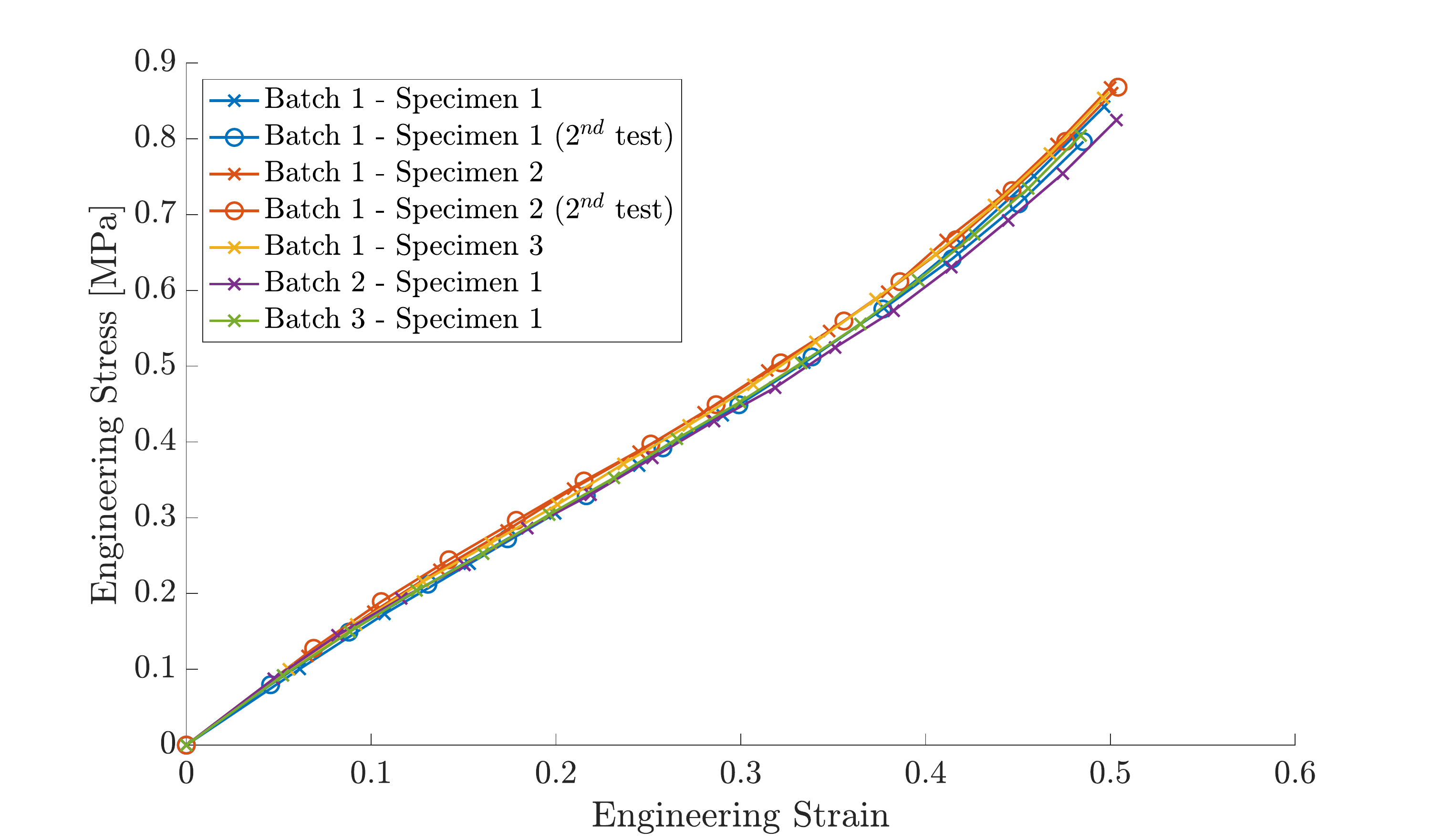}
	\caption{\rev{Engineering stress plotted against engineering strain for seven tensile tests performed on dog-bone specimens of PDMS. Tests repeated on the same sample are plotted in the same color.}}
	\label{fig4:tensile}
\end{figure}

Dog-bone specimens, with a gauge length of $12$~mm and a width of $2.2$~mm, are punched out of PDMS slabs, which are processed along with the manufacturing of the metamaterial specimens. 
The thickness of these dog-bone specimens varies between $1.0$ to $1.5$~mm for different batches.
\rev{The specimens are made to resemble those of the ASTM D412-C standard, but with smaller dimensions.}
A similar speckle pattern as described in Section \ref{sec4:dic} is applied on the dog-bone samples, and \textit{in-situ} tensile tests are performed.
Loads are measured with a $50$~N load cell.
Global DIC, in which the displacement field is parametrized with linear polynomials, is used to measure the strains inside the specimens.
\rev{ Note that since the strains are measured using DIC, small deviations from the ASTM standard introduce no inaccuracies in the measurements.
Tensile tests are repeated on five specimens produced from three different batches of PDMS processing. 
Two of the specimens are tested twice.
The results of all the seven tests are shown in Fig.~\ref{fig4:tensile}, where engineering stress is plotted against engineering strain, exhibiting the good repeatability of the tensile tests, as well as the repeatability of the PDMS processing. 

Adopting the hyper-elastic Ogden material model, the material parameters are identified by minimizing the difference between the reaction force of a finite element model in a uniform state of deformation subjected to uniaxial tension and the experimentally measured forces of one of the tensile tests.}
The strain energy density function is given by:
\begin{equation}
W = \frac{1}{2}\kappa \left(J-1\right)^{2}+\sum^{N}_{k=1} \frac{c_k}{m_k^2}\left(\lambda_1^{m_k}+\lambda_2^{m_k}+\lambda_3^{m_k}-3-m_k \ln J\right),
\end{equation}
where $\lambda_i$ are the principal stretches, $J = \det(\boldsymbol{F})$ is the volume change ratio, $\boldsymbol{F}$ the deformation gradient tensor, $\kappa$ is the bulk modulus, and $c_k$ and $m_k$ are the material parameters, with $N$ taken equal to 2 in this study \citep{Vossen2015}.
The principal stresses, $\sigma_i$, are given as:
\begin{equation}
\sigma_i = \kappa \left(J-1\right) + \sum^{N}_{k=1} \frac{1}{J} \frac{c_k}{m_k}\left(\lambda_i^{m_k}-1 \right),
\end{equation}

The initial values for $c_k$ and $\lambda_k$ as well as the bulk modulus are chosen from \cite{Gerratt2013}, due to the similar PDMS processing  utilized in that study.
Considering the variety of values reported in the literature for the bulk modulus of Sylgard 184, varying from  20 to $1200$~MPa \citep{Vossen2015,Gerratt2013,Kim2011}, it was decided to perform confined compression tests on circular samples of $5$~mm diameter punched from PDMS slabs to evaluate the bulk modulus. 
An average of $415$~MPa and standard deviation of $125$~MPa was found for bulk modulus over ten tests.
The low average is due to air bubbles observed under optical microscope, while the large standard deviation is due to the non-uniform distribution of the air bubbles.
It is realized, however, that  the degassing process in the metamaterial specimens is less effective, considering their thickness and complicated design, which will for sure leave more bubbles in these metamaterial specimens than for the flat compression specimens and thus results in a lower bulk modulus.
Therefore, the low initial guess for the bulk modulus obtained from Ref.\ \citep{Gerratt2013}, i.e.\ $29.15$~MPa, seems to be a reasonable value to use as initial guess.

A staggered approach is used to identify all five material parameters.
First, $c_k$ are optimized for strains up to 2\%, then $c_k$ and $m_k$ are optimized for the entire strain range, i.e.\ up to 60\%.
The bulk modulus is finally found for the whole strain range while keeping the optimized values of $c_k$ and $m_k$ constant.
This approach results in optimized values of $\kappa = 29.6716$~MPa, $c_1 = 0.0892$~MPa, $c_2 = 1.2537$~MPa, $m_1 = 10.0959$ and $m_2 = 0.0036$.
Note that a separate numerical investigation revealed only 2\% change in initial mechanical response  of metamaterial specimens for a change of bulk modulus from 30  to $1000$~MPa, thus revealing a very small sensitivity to this material parameter.

\section*{Data availability}
The raw/processed data required to reproduce these findings can not yet be shared at this time as the data forms part of an ongoing study.

%
%
\bibliography{bibfile_Thesis}
\end{document}